\documentclass[final,3p,times,twocolumn]{elsarticle}
\def \Am2{\AA$^{-2}$}
\def \ph2{{\em p}-H$_2$}
\def \he4{$^4$He}
\usepackage{amssymb}
\journal{Results in Physics}

\begin{document}

\begin{frontmatter}



\title{Quantum Monte Carlo study of thin parahydrogen films on graphite}

\author{Jie-Ru Hu\fnref{pipo}}
\author{Massimo Boninsegni\corref{pupo}}
\cortext[pupo]{Corresponding author: m.boninsegni@ualberta.ca}
\fntext[pipo]{Present address: State Key Laboratory of Precision Spectroscopy, East China Normal University, Shanghai 200062, China.}

\affiliation{organization={Department of Physics},%Department and Organization
            addressline={University of Alberta}, 
            city={Edmonton},
            postcode={T6G 2H5}, 
            state={Alberta},
            country={Canada}}

\begin{abstract}
The low-temperature properties of one and two layers of parahydrogen adsorbed on graphite are investigated theoretically through Quantum Monte Carlo simulations. We adopt a microscopic model that explicitly includes the corrugation of the substrate. We study the phase diagram of a monolayer up to second layer promotion, and the possible occurrence of superfluidity in the second layer. We obtain results down to a temperature as low as 8 mK. We find second-layer promotion to occur at a considerably greater coverage than obtained in previous calculations and estimated experimentally; moreover, we find no evidence of a possible finite superfluid response in the second layer, disproving recent theoretical predictions.

\end{abstract}


\end{frontmatter}


\section{Introduction}\label{intro}
The physics of a thin (few layers) film of parahydrogen molecules (\ph2) adsorbed on a strongly attractive corrugated substrate like graphite, has been extensively studied experimentally \cite{Nielsen1980,Lauter1990,Wiechert1991,Wiechert1991b,Morton2000} and theoretically \cite{Wang1980,Ni1986,Novaco1988,Gottlieb1990,Nho2002,soliti2022}. The basic features of the phase diagram of the system at low temperature are well understood, but there exist specific aspects which still require elucidation. 
\\ \indent
Adsorption of \ph2  on essentially {\em any} substrate \cite{Boninsegni2004} occurs through the formation of successive crystalline layers. On a strongly attractive substrate such as graphite, the first layer is physically distinct, featuring an equilibrium phase registered with the underlying carbon lattice, which transitions at higher coverage to a compressible, incommensurate crystal through a series of domain-wall phases.  It is interesting to compare \ph2   with \he4, i.e., the archetypal (Bose) quantum fluid, and it seems fair to state that there is little qualitative difference between the phase diagram of the first layer of either substance on graphite, as the energetics is in both cases dominated by the substrate attraction. On the other hand, the second layer displays very different behavior for the two systems, with \he4 forming a liquid (superfluid) film while \ph2 remains solid.
\\ \indent 
An outstanding puzzle is represented by the value of coverage $\theta_p$ of a \ph2  monolayer at which molecular promotion to the second layer begins to occur. There exists an experimental estimate \cite{Wiechert1991} for $\theta_p \approx 0.094$ \Am2, considerably {\em lower} (by about 15\%) than the corresponding coverage for a \he4 monolayer on the same substrate \cite{Greywall1991}. This seems counter-intuitive, as {\em a}) one would expect a parahydrogen monolayer to be more compressible than one of helium, which displays much more pronounced quantum effects \cite{notea}, and {\em b})  the attractive well of the effective potential between a \ph2 molecule and a graphite substrate \cite{Crowell1982} is about three times deeper than that for a \he4 atom and the same substrate \cite{Carlos1980}. 
\\ \indent 
Microscopic calculations for realistic models of the system have yielded conflicting results; specifically, an early study \cite{Nho2002} making use of Quantum Monte Carlo (QMC) simulations predicted second-layer promotion at a coverage very close to that reported in Ref. \cite{Wiechert1991}, but a subsequent study \cite{Dusseault2018} of the same system based on a slightly different QMC methodology showed no promotion to second layer up to a coverage $\sim 0.110$ \Am2, i.e., very similar to that observed for a \he4 monolayer. Both calculations treated the underlying graphite substrate as smooth, i.e., ignored its corrugation. While the cause of the disagreement between the two calculations is unclear, it is conceivable that corrugation could have a significant effect on $\theta_p$. However, in Ref. \cite{Dusseault2018} the same calculation was carried out for a \ph2 monolayer on graphene (which is some 10\% less attractive than graphite), explicitly including substrate corrugation, and the second layer promotion coverage was found to be close to 0.11 \Am2. One certainly expects a comparable or even greater value for the more attractive graphite. 
\\ \indent
A second point of contention has to do with the second (solid) layer, for which the claim of a possible finite superfluid response in the ground state has been recently made \cite{soliti2022}. This seems very similar to other predictions of superfluidity (SF) of either solid helium or hydrogen in various settings (usually in confinement) arrived at using the same ground state methodology, all of which have been shown to be incorrect. There is presently no experimental evidence of superfluid behavior of \ph2, except (possibly) in very small clusters ($\sim$ 15 molecules or less) \cite{Grebenev2000}; all reliable theoretical studies of parahydrogen, including in reduced dimensions \cite{Boninsegni2013}, thin films \cite{Boninsegni2021} and in disordered and/or inhomogeneous environments \cite{Turnbull2008,Boninsegni2018,Boninsegni2016} have shown that exchanges of indistinguishable molecules, which are known to underlie SF, are strongly suppressed even at low temperature. However, no microscopic calculation for the second layer of \ph2 on graphite has yet been carried out, independent of that of Ref. \cite{soliti2022}, in which the corrugation of the graphite substrate is taken into account. Although a recent study \cite{Boninsegni2023} for \he4 provides strong evidence that the effect of substrate corrugation on the superfluid properties of the second layer is minimal, it seems worthwhile to investigate in greater depth the scenario of a possible superfluid response on the second layer of \ph2 on graphite, taking the corrugation of the substrate into account. 
\\ \indent
To clarify the above outstanding issues, we have carried out extensive QMC simulations of a thin (1 and 2 layers) film of \ph2 adsorbed on graphite at low temperature (as low as 8 mK), making use of the same  
{\em ab initio} potential proposed in Ref. \cite{Nho2002}, describing the interaction of a \ph2 molecule with the substrate, specifically designed to capture the effects of substrate corrugation. We aim to compare directly our results to those obtained therein.
For the 2-layer system, we focused on ascertaining whether there is a significant enhancement of quantum exchanges of indistinguishable \ph2 molecules at subKelvin temperatures, and with that the possible onset of a weak superfluid response as predicted in Ref. \cite{soliti2022}, at the same thermodynamic conditions.
\\ \indent
Our results yield, for the monolayer system, a second-layer promotion coverage close to 0.110 \Am2, i.e., consistent with both that found on graphene as well as on a smooth graphite substrate  \cite{Dusseault2018}, significantly above the existing experimental estimate from Ref. \cite{Wiechert1991}. While our results are generally {\em qualitatively} consistent with those of Ref. \cite{Nho2002}, there are {\em significant} numerical discrepancies, which we believe may be at the root of the difference between the promotion coverage predicted therein and in this work. Still, the disaccord of our calculation with experiment remains, and at this point can only be solved by new, independent experimental and theoretical studies.
\\ \indent
{\em No evidence} of any ``supersolid'' phase is observed for the second layer. Our simulations, carried out down to a temperature as low as 8 mK, yield the same paucity of quantum-mechanical exchanges of indistinguishable molecules observed at temperatures of the order of 1 K,  down to temperatures more than two orders of magnitude lower; no discernible difference can be seen in the one-body density matrix computed at temperature $T=4$ K and $T=$0.03 K, once again confirming that this system forms an insulating crystal which does {\em not} undergo a superfluid transition in the $T\to 0$, which is the conclusion reached in many previous theoretical studies, consistent with all existing experimental observations. It is therefore our submission that the prediction of Ref. \cite{soliti2022} is incorrect. 
\\ 
\indent
The remainder of this paper is organized as follows: in section \ref{mod} we describe the microscopic model of the system; in Sec. \ref{meth} we briefly describe our methodology; we present and discuss our results in Sec. \ref{res} and finally outline our conclusions in Sec. \ref{conc}.
\section{Model}\label{mod}
We consider an ensemble of $N$ \ph2 molecules, regarded as point-like spin-zero bosons, moving in the presence of a graphite substrate, modeled as described below. 
The system is enclosed in a simulation cell of sizes $L_x\times L_y\times L_z$,  with periodic boundary conditions in all directions (but $L_z$ is taken large enough to make boundary conditions in the $z$ direction irrelevant). The graphite substrate lies on the $z=0$ plane; in our monolayer simulations, we set $L_x=34.08$ \AA, $L_y=36.8927$ \AA; on the other hand, those of a two-layer system were carried out for a significantly {\em smaller} simulation cell, one with $L_x=21.3$ \AA\ and $L_y=19.6761$ \AA. The reason is that our goal in this case is exclusively that of ascertaining whether the conditions for SF exist, and the superfluid response is typically enhanced in a finite system with periodic boundary conditions. Thus, if no superfluid signal is observed in a system of small size, {\em a fortiori} that will be the case in the thermodynamic limit.
\\ \indent
The quantum-mechanical many-body Hamiltonian reads as follows:
\begin{eqnarray}\label{u}
\hat H = -\lambda\sum_{i}\nabla^2_{i}+\sum_{i<j}v(r_{ij})+\sum_{i}V({\bf r}_{i}).
\end{eqnarray}
The first and third sums run over all the $N$ \ph2 molecules,  $\lambda=12.031$ K\AA$^{2}$; the second sum runs over all pairs of molecules, $r_{ij}\equiv |{\bf r}_i-{\bf r}_j|$, 
${\bf r}_i\equiv(x_i,y_i,z_i)$ being the position of the $i$th molecule, and $v(r)$ is the accepted Silvera-Goldman pair potential \cite{Silvera1978}, which describes the interaction between two \ph2 molecules. 
$V$ is the potential describing the interaction of a \ph2 molecule with the graphite substrate; for consistency with the calculation of Ref. \cite{Nho2002}, we use the same potential utilized therein, which can be expressed as follows:
\begin{equation}\label{carloscole}
    V({\bf r}) = V_0(z) + \sum_{\bf G} V_{\bf G}(z)\ {\rm exp}\  [ i\ (G_xx+G_yy)]\ 
\end{equation}
$V_0(z)$ is a term that only depends on the distance of the molecule from the basal plane; it corresponds to the laterally averaged potential normally utilized to represent a flat substrate \cite{Crowell1982}.  The sum in the second term of the right-hand side of (\ref{carloscole}) runs over all reciprocal lattice vectors ${\bf G}\equiv (G_x,G_y,0)$ of the graphite substrate. The functions $V_0(z), V_{\bf G}(z)$ are provided in Ref. \cite{Nho2002}. We come back to the details of the evaluation of (\ref{carloscole}) in our simulations in Sec. \ref{meth}. 
%\begin{figure}[h]
%\centering
%\includegraphics[width=\linewidth]{scheme.pdf}
%\caption{Schematic of the simulation setup. The simulation  cell is elongated, sandwiched between %homogeneous solid and superfluid phases,
%modeled as homogeneous, continuous media.}
%\label{scheme}
%\end{figure}
\section{Methodology}\label {meth}
The QMC methodology adopted here is the canonical \cite{Mezzacapo2006,Mezzacapo2007} version of the continuous-space Worm Algorithm \cite{Boninsegni2006,Boninsegni2006b}, a well-established finite temperature QMC technique. 
Details of the simulations carried out in this work are standard, and therefore the reader is referred to the original references. However, because as mentioned above there are numerical discrepancies between our results and those reported in Ref. \cite{Nho2002}, where a similar computational technique was utilized, we are going to provide here enough technical information to enable others to repeat the calculation, in order to investigate the possible causes of the disagreement.
\\ \indent
The calculation of the potential energy of interaction between a \ph2 molecule and the graphite substrate (Eq. \ref{carloscole}) has been carried out in the simulation by tabulating and interpolating the functions $V_0(z)$ and $V_{\bf G}(z)$, and by performing a direct, on-the-fly summation over terms associated to a subset of reciprocal lattice vectors; we found that the inclusion of the twelve shortest reciprocal lattice vectors is sufficient to achieve a numerically accurate representation of $V({\bf r})$. 
\\ \indent
The key physical quantities computed in this work are the energetics and the superfluid fraction $\rho_S(T)$ of the top layer as a function of temperature, for which we use the well-known winding number estimator \cite{Pollock1987}. We also evaluate structural properties, such as the averaged molecular density profile along the direction $z$,  perpendicular to the substrate, in order to investigate the occurrence of promotion to the second layer.  Crystalline order in both layers may also be monitored through the visual inspection of the imaginary time paths. 
\\ \indent
Simulations of a two-layer system have been carried out by {\em a}) considering molecules in the top and bottom layer as two distinct species and {\em b}) regarding molecules in the bottom layer as {\em distinguishable} quantum particles (i.e., ``Boltzmannons''). Both approximations are justified by the {\em de facto} absence of quantum-mechanical exchanges involving molecules in the bottom layer; on the other hand, molecules in the top layers are considered indistinguishable and their quantum (Bose) statistics is fully accounted for (as we shall see, even changes of molecules in the top layers turn out to be exceedingly infrequent even at the lowest temperature considered here). 
\\ \indent
We used the primitive approximation for the short imaginary-time ($\tau$) propagator and carried out extrapolation of all the computed physical observable in the $\tau\to 0$ limit. This point warrants a quantitative discussion because this is where the first major difference with Ref. \cite{Nho2002} is noted.
\begin{figure}[h]
\centering
\includegraphics[width=\linewidth]{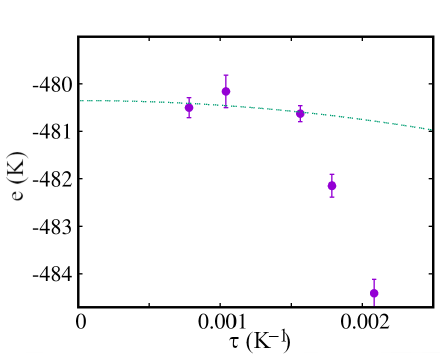}
\caption{Energy per molecule at $T=1$ K, for coverage $\theta=0.0636$ \Am2, computed as a function of the time step $\tau$. Dotted line is a quadratic fit to the data for $\tau\to 0$. It is not possible to include in the quadratic fit data points obtained for values of the time step above $1.5625\times 10^{-3}$ K$^{-1}$.}
\label{extra}
\end{figure}

Fig. \ref{extra} show the computed energy per \ph2 molecule at temperature $T=1$ K as a function of the time step $\tau$, for coverage $\theta=0.0636$ \Am2, namely the coverage at which a commensurate monolayer crystal forms, as will be shown below. Because the primitive approximation is used, one expects quadratic behavior as a function of $\tau$, in the $\tau\to 0$ limit.  As shown in the figure, the estimate for $\tau=1.5625\times 10^{-3}$ K$^{-1}$ is indistinguishable, within statistical uncertainties, from the extrapolated one, namely $-480.3(4)$ K. This is only a few K lower than the value reported in Ref. \cite{Nho2002}, namely $-476.9$ K, but what is puzzling is that the value of the time step used therein was 0.005 K$^{-1}$, which in our calculation yields an energy estimate some $\sim 17$ K {\em lower}  than the value reported in Ref. \cite{Nho2002}. The reason for this disagreement is unknown to us, at this time, as the calculations are based on the same potentials; it cannot, in our view, be attributed to the different system sizes (the calculation of Ref. \cite{Nho2002} used $N=60$ molecules, whereas $N=80$ was the number in this work, at this coverage), nor to the difference in temperature between the two calculations (we come back to this point below), nor to a possible, different choice of wave vectors included in the expansion of the potential (\ref{carloscole}). 
\\ \indent
The energy estimates quoted below for the energy were all obtained either by carrying out explicitly the $\tau\to 0$ extrapolation, as shown above, or using a time step  $\tau=1.5625\times 10^{-3}$ K$^{-1}$. On the other hand, it has been observed in this work, as well as in \he4 studies based on the same methodology \cite{Boninsegni2023}, that convergence of the results for structural and superfluid properties is achieved with a considerably greater (up to four times) time step than that needed for the energy. 
\section{Results}\label{res}
\subsection{Monolayer}
We begin by discussing the energetics of a monolayer. In general, we observe that the estimates for the energy, as well as for all structural properties, remain essentially unchanged (the energy within less than 1\%) for temperatures below 4 K.
In the zero-coverage limit (i.e., one particle), we obtain an energy per molecule of $-454.7(1)$ K, which is very close to the estimate obtained by Crowell and Brown for the laterally averaged potential \cite{Crowell1982}.
\begin{figure}[h]
\centering
\includegraphics[width=\linewidth]{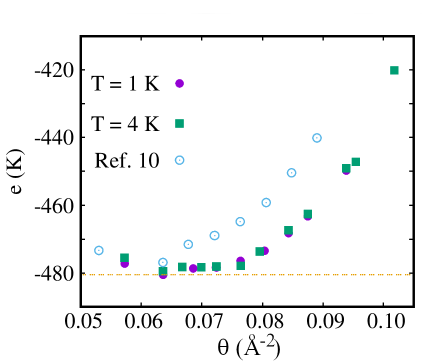}
\caption{Energy per \ph2 molecule on a corrugated graphite substrate computed in this work as a function of coverage $\theta$, at the two temperatures $T=1$ K (filled circles) and $T=4$ K (filled boxes). Dashed line is the ground state energy at the equilibrium density,  attained for coverage $\theta=0.0636$ \Am2. Open circles refer to data from Ref. \cite{Nho2002} at $T=2$ K. Statistical errors are smaller than symbol sizes.}
\label{enT1}
\end{figure}

In our monolayer calculations, we cut off the pair potential at 17 \AA, i.e., we set it to zero beyond that distance. We estimate the energy contribution arising from pairs of molecules at distances greater than that to amount to less than 0.01\% of the total energy.\\
Fig. \ref{enT1} shows the computed energy per \ph2 molecule as a function of coverage, at the two temperatures $T=1$ K (filled circles) and $T=4$ K (filled boxes). Also shown (open circles) are the estimates from Ref. \cite{Nho2002}. There is an obvious numerical discrepancy between the results obtained here, which on the scale of the figure are temperature independent, and those offered in Ref. \cite{Nho2002}. The difference is just a few K at the equilibrium coverage $\theta=0.0636$ \Am2, which is the same on both works and corresponds to the formation of a solid monolayer commensurate with the underlying substrate, but increases at higher coverage and becomes as large as $\sim 20$ K in the proximity of the coverage estimated in Ref.
\cite{Nho2002} to correspond to second layer promotion.
\begin{figure}[h]
\centering
\includegraphics[width=\linewidth]{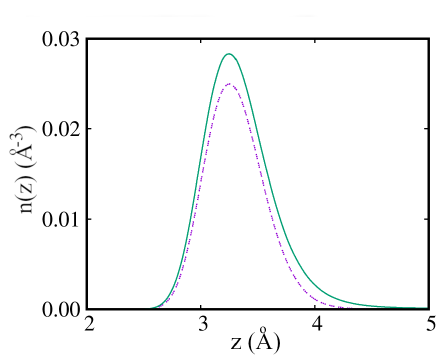}
\caption{Density profile $n(z)$ (in \AA$^{-3}$) in the direction perpendicular to the graphite substrate for a monolayer \ph2 film of coverage $\theta=0.1098$ \Am2, computed by simulation at temperature $T=1$ K. Dashed line shows the corresponding profile for coverage $\theta=0.0636$ \Am2.}
\label{profiles}
\end{figure}

The origin of such a large disagreement between these two calculations is unclear, especially given that, as shown above, the choice of time step made by the authors of Ref. \cite{Nho2002} should have resulted in an {\em underestimation} of the energy. While the overall trend is similar in both calculations, the specific physical predictions made in Ref. \cite{Nho2002} are, in our view, likely to be affected by the quantitative inaccuracy of their energy estimates.
\\ \indent
Altogether, the physics of the monolayer is very similar to that observed on graphene; we did not attempt to characterize it in detail in this work, focusing instead on the issue of second-layer promotion, attempting to elucidate the outstanding disagreement between some of the theoretical calculations and the existing experimental estimates. 
\\ \indent
Fig. \ref{profiles} shows the \ph2\  density profile $n(z)$ in the direction perpendicular to the substrate, computed at $T=1$ K for a coverage $\theta=0.1098$ \Am2, i.e., well above the experimentally estimated second-layer promotion coverage $\theta_p$ \cite{Wiechert1991}. 
\begin{figure}[h]
\centering
\includegraphics[width=\linewidth]{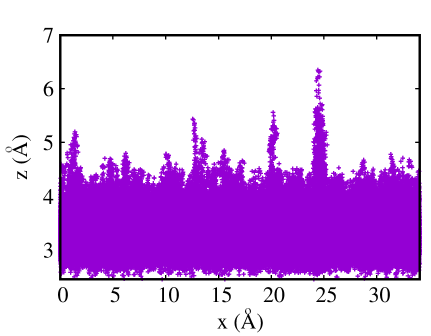}
\caption{Snapshot of instantaneous many-particle world lines for a \ph2 monolayer of coverage $\theta=0.1016$ \Am2 at temperature $T=1$ K, projected on the $x-z$ plane.}
\label{flares}
\end{figure}
In Ref. \cite{Nho2002} it was argued that simulations based on the many-body Hamiltonian (\ref{u}) yield a value of the second-layer promotion coverage in agreement with experiment, based on an analysis of density profiles such as that shown in Fig. \ref{profiles}, displaying a ``bump'' at distances $\sim 6.5$ \AA\ from the substrate for coverages close to, or above the experimentally estimated $\theta_p$. Our simulations, on the other hand, give no evidence of that until a much higher value of coverage is reached, slightly above 0.11 \Am2, which is consistent both with the value found on graphene \cite{Dusseault2018} as well as for \he4 on graphite \cite{Corboz2008}. 
\\ \indent
Fig. \ref{flares} shows an instantaneous configurational snapshot (particle world lines) from a simulation of a \ph2 monolayer at temperature $T=1$ K for coverage $\theta=0.1018$ \Am2. Although the world lines extend far out, up to distances considerably greater than the average width of the layer, nevertheless all particles remain localized within it, as shown by the positions of the path centroids. There is no evidence of the formation of an actual second layer. We therefore conclude that promotion to the second layer does not take place for $\theta\sim 0.094$ \Am2 as contended in Ref. \cite {Nho2002}, but at considerably higher coverage, $\sim 0.11$ \Am2, value is very close to that predicted for \he4 as well, on the same substrate \cite{Corboz2008}.
\subsubsection{Bilayer}
We now discuss the results of our simulation of a two-layer \ph2 system, aimed at investigating the possible occurrence of a finite superfluid response in the top layer, in the low-temperature limit. The possible SF of a film of \ph2 has been investigated and ruled out in numerous theoretical calculations, characterized by a wide variety of physical settings \cite{Boninsegni2021,Turnbull2008,Boninsegni2018,Boninsegni2016}. Every time the conclusion was arrived at that exchanges among identical molecules, which underlie any superfluid response, are too strongly suppressed in this system, mainly due to the large value of the diameter of the repulsive core of the potential at short distances.

\begin{figure}[h]
\centering
\includegraphics[width=\linewidth]{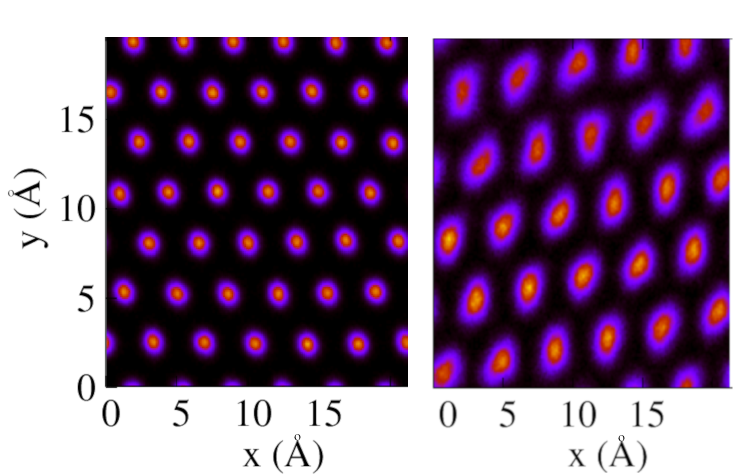}
\caption{Density map of \ph2 molecules in the bottom (left) and top (layer), at temperature $T=8$ mK. The total coverage is 0.165 \Am2, the two-dimensional density of the bottom layer is 0.100 \Am2.}
\label{map}
\end{figure}

Nonetheless, it was recently contended that the top layer of a two-layer film of coverage 0.165 \Am2 may display a small but finite superfluid response in the ground state \cite{soliti2022}. If confirmed, this would be a remarkable finding, as the top layer still displays crystalline long-range order, which would render this the first example of a {\em supersolid} system (there is no indication that the top solid layer would be commensurate with the underlying one).
\\ \indent
In order to provide an independent theoretical assessment of this intriguing prediction, we carried out simulations at finite temperature of the same system, based on model (\ref{u}), which is physically equivalent to that utilized in Ref. \cite{soliti2022}, as explicitly verified in Ref. \cite{Boninsegni2023}. As mentioned above, we purposefully utilized a relatively small simulation cell and overall number of particles, enabling us to explore as wide a range of temperature as possible, enabling us to detect any sign of build-up of superfluid order. We set the coverage to be the same as that of Ref. \cite{soliti2022}, namely 0.165 \Am2, with a two-dimensional density equal to 0.100 \Am2\ in the bottom layer \cite{note3}, and 0.065 in the top one, just as in Ref. \cite{soliti2022}. It is worth restating that the absence of SF in such a small system which comprises 42 \ph2 molecules on the bottom layer and 27 on the top one, {\em necessarily} implies no SF in the thermodynamic limit.
\\ \indent
Fig. \ref{map} shows two-dimensional density maps for the bottom and top layers of the simulated two-layer system, at a temperature $T$=8 mK. The first obvious remark is that the system forms crystal phases on both layers, as expected; \ph2 molecules are initially given the positions shown in the figure for the bottom layer, namely a triangular lattice incommensurate with the graphite substrate is assumed, consistently with our present experimental and theoretical understanding of the system. Molecules in the top layer, on the other hand, are initially placed at {\em random} initial position at a distance $z=6$ \AA\ from the graphite basal plane, i.e., the regular crystalline arrangement shown in the right part of Fig. \ref{map} forms {\em spontaneously}, despite the fact that the cell is not designed to accommodate a triangular lattice with that number particles; just as observed in other studies, if the system cannot form the preferred structure, due to geometrical constraints, it will do the ``next best thing'', namely form whatever triangular crystal can be formed inside the cell.
\\ \indent
Another thing that can be noticed is that \ph2 molecules in both layers are highly localized, that there is no visible overlap between quantum delocalization clouds of adjacent molecules to indicate that exchanges among molecules are essentially non-existent, and this is precisely what is observed in the simulations, i.e., even those carried out at a temperature as low as 8 mK show the nearly complete {\em absence} of exchanges of \ph2 molecules. Consequently, no evidence of any finite superfluid response is observed; the superfluid fraction of the top layer is {\em zero} (within the precision of the computing machine) even at a temperature as low as $T=8$ mK. This observation is consistent with all other studies of \ph2 performed by us, though it is worth mentioning that, at least to our knowledge 8 mK is the lowest temperature attained so far in QMC simulations of either \he4 or \ph2.
\begin{figure}[h]
\centering
\includegraphics[width=\linewidth]{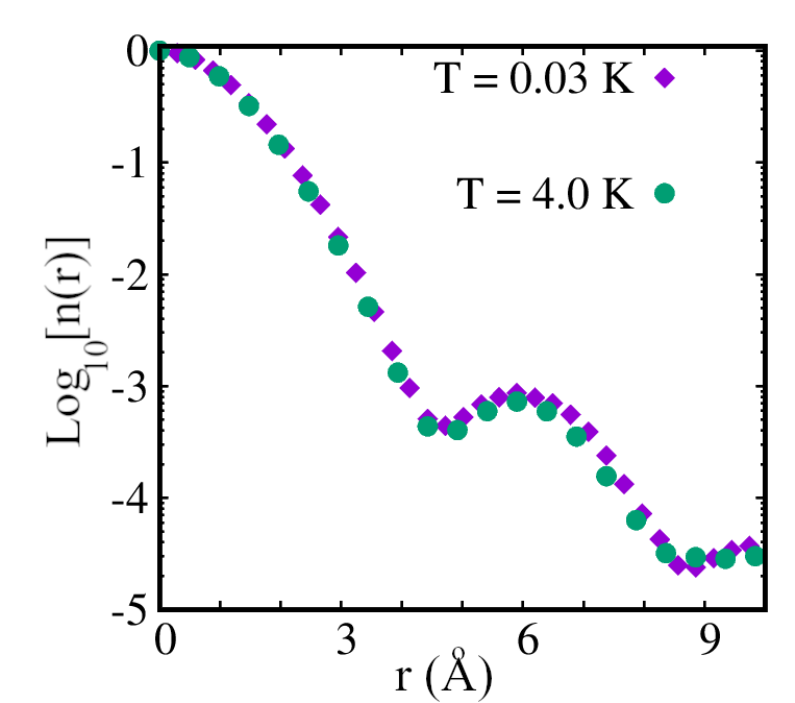}
\caption{Laterally averaged one-particle density matrix for the top layer of \ph2 in a two-layer system of coverage 0.165 \Am2. Results shown are for the two temperature $T=4$ K (circles) and
$T=0.03$ K (diamonds). Statistical errors are smaller of symbol sizes.}
\label{nofr}
\end{figure}

Fig. \ref{nofr} compares the laterally averaged one-body density matrix computed for the top layer, at the two temperatures $T=4$ K and $T=0.03$ K. The most remarkable aspect of this result, to our knowledge, is the virtual {\em indistinguishability} of the two results, at least within the scale of the figure. The absence of any noticeable temperature dependence within a temperature range that spans over two orders of magnitude constitutes, in our view, strong evidence that the system is {\em not} superfluid, even in the $T=0$ limit. One could argue, of course, that the temperature that we considered is not low enough to observe the onset of SF. We do not believe that to be the case, however, as the hypothetical superfluid transition should conform to the Bereszinskii-Kosterlitz-Thouless paradigm, and in particular, fulfill the well-known ``universal jump'' condition \cite{Nelson1977}, based on which one can obtain a fairly reliable quantitative estimate \cite{Zhang2023} of the superfluid transition temperature $T_c$. In this case, we can expect $T_c\approx$ 10 mK, i.e., simulations of a system of size as small as that considered here at $T=8$ mK should yield clear evidence of quantum-mechanical exchanges of top-layer molecules and/or a finite in-plane superfluid response, if the predictions of Ref. \cite{soliti2022} were correct. 

\section{Discussion and Conclusions}\label{conc}
We have carried out extensive QMC simulations of monolayer and bilayer \ph2 films adsorbed on a graphite substrate, making use of a microscopic model in which the corrugation of the substrate is fully taken into account. For a monolayer, we report important numerical differences with respect to prior theoretical work \cite{Nho2002}, the most important being the estimated second-layer promotion coverage, which we find to be some $\sim$ 15\% higher. The resolution of the discrepancy requires additional independent theoretical studies carried out with the same methodology and microscopic model. For the moment, however, we note that significantly greater binding energy of the \ph2 molecules obtained in this work, at coverages above the equilibrium one. Our estimate is also in disagreement with the only (at least to our knowledge) existing experimental determination of the second-layer promotion coverage for this system; while we cannot comment on experimental details, it seems counterintuitive that second-layer promotion should occur at a {\em lower} coverage for a system such as \ph2, which experiences a significantly stronger attraction to the graphite substrate and whose behavior is considerably less quantum than \he4. Nonetheless, it should be mentioned that our result is also at variance with that of a recent ground state calculation \cite{soliti2022}, based on a different model of the substrate, which predicts a value of the second layer promotion coverage in agreement with the experiment.
\\ \indent
We have also investigated the possible occurrence of a finite superfluid response in the top layer of a bilayer system, at the same condition of coverage in which other authors have maintained that SF could occur in the low-temperature limit. Our simulations, reaching a temperature as low as 8 mK, confirm the same observations consistently and repeatedly made for \ph2 in all physical settings explored so far, namely that this system fails to display SF due to the strong suppression of quantum-mechanical exchanges. This parallels the conclusion reached for $^4$He monolayers on graphene, for which a very similar contention of SF was made, based on the same approach \cite{Happacher2013}.
In our submission, our calculation {\em disproves} the contention of Ref. \cite{soliti2022}, the latest of a series of incorrect predictions \cite{Boninsegni2013,Boninsegni2018,Boninsegni2016,Boninsegni2011,Cinti2019} of a finite SF response based on a computational technology (Diffusion Monte Carlo) known to be affected by serious limitations, chiefly the bias due to the use of a finite population of random walkers \cite{Boninsegni2001,Boninsegni2012b,Inack2018,Ghanem2021,Brand2022}. 
\\ \indent
In general, all bulk supersolid phases that have been established theoretically hinge on some ``softening'' of the repulsive core of the pairwise interaction at short distances \cite{Boninsegni2012c,Kora2019}.

\section{Declaration of Competing Interests}
The authors declare no known competing financial interests or personal relationships that could appear to influence the work reported in this paper.

\section{Acknowledgments}
This work was supported by the Natural Sciences and Engineering Research Council of Canada.

 \bibliographystyle{elsarticle-num} 
 \bibliography{refs}

\begin{thebibliography}{10}
\bibitem{notea}
For example, the equilibrium density of \ph2 is almost 50\% greater than that of \he4 in two dimensions.
\end{thebibliography}





% \end{thebibliography}
\end{document}